\documentclass[prb,aps,superscriptaddress,amssymb,floatfix,showpacs,amsfonts,eqsecnum]{revtex4} 
\usepackage{graphicx,psfrag,amsmath,amssymb,float,subfigure}
\usepackage{color}
\input{epsf}
\usepackage{graphicx,natbib}

\newcommand{\bc}{\begin{center}}
\newcommand{\ec}{\end{center}}
\newcommand{\be}{\begin{equation}}
\newcommand{\ee}{\end{equation}}
\newcommand{\bea}{\begin{eqnarray}}
\newcommand{\eea}{\end{eqnarray}}
\newcommand{\nn}{\nonumber \\}

\def\12{\frac{1}{2}}

\makeatletter
\usepackage{psfrag}

\begin{document}

\title{Vortex interactions in a thin platelet superconductor}

\author{Cedric Yen-Yu Lin} 
\author{Ian Affleck}

\affiliation{Department of Physics and Astronomy, University of British Columbia,
Vancouver, BC, Canada, V6T 1Z1}

\date{\today}

\begin{abstract}
The thermal fluctations of vortices in a superconductor  
can be usefully mapped onto the quantum fluctations of a collection of bosons at $T=0$ moving in
 2 dimensions. 
When the superconductor is a thin platelet with the magnetic field parallel to its surface, the interacting 
quantum bosons are effectively moving in 1 dimension, allowing for powerful Luttinger liquid 
methods to be applied. Here we consider how this 1 dimensional limit is approached, 
studying the interaction of vortices with the platelet surfaces 
and each other.
Using realistic parameters and vortex interactions for an underdoped YBCO platelet
we determine the scattering length, $a$, characterizing the low energy interaction of a vortex pair as 
a function of the platelet thickness.  
$a$ determines the Luttinger parameter, $g$, for the quantum system at low densities, $n_0$: $g\to 1-2an_0$.

\end{abstract}


\maketitle
\section{Introduction}
Thermal fluctuations of vortices, taking into account pinning by impurities and vortex-vortex interactions, is a challenging 
and technologically important problem in statistical physics. An elegant approach to this subject is to 
map each fluctating vortex line into the world line of a quantum particle in a Feynman path integral, with 
the magnetic field direction becoming the imaginary time direction and the bosons moving 
in the other two spatial directions.\cite{Nelson,Fisher} Such a mapping is especially powerful 
for studying columnar defects which become static point defects in the quantum model. When the 
field direction is tilted relative to the (parallel) pins a novel non-Hermitean 2-dimensional many-body quantum problem arises.\cite{Nelson2} 
If the superconductor is a thin platelet, of thickness of order the penetration depth, with the field lying in the plane, 
then the quantum bosons are essentially restricted to one dimension.  This allows theoretical techniques 
including Tomonoga Luttinger liquid (TLL) theory and Density Matrix Renormalization Group (DMRG) to be brought to 
bear,\cite{Hofstetter,Affleck} 
rendering tractible a formidable problem.  It may be feasible to realize this classical analogue of a Luttinger liquid experimentally 
using high-T$_c$ superconductors. A promising candidate would be a very clean highly underdoped YBCO single 
crystal in which the penetration depth $\lambda_c$ can be as large as 50 microns.  
A platelet should be cleaved with thickness in the $a$-direction of order .1  to 1 mm. and then a magnetic 
field should be applied in the $b$-direction.

It was shown in [\onlinecite{Hofstetter,Affleck}] that critical phenomena connected with rotating 
the field direction away from the pin direction is controlled by the dimensionless Luttinger parameter, $g$. 
 In the case $g>1$ columnar 
defects are irrelevant and have little effect on the long-distance properties of the vortices. 
On the other hand, for $g<1$ they are relevant and an arbitrarily weak pinning potential 
drastically alters the system. Thus it is of considerable interest to determine $g$ 
and how it depends on the parameters of the system, including the density, $n_0$. In the dilute limit 
$g$ approaches unity and the interacting boson system becomes equivalent to non-interacting fermions. 
The leading density dependent correction is\cite{Affleck}
\be g=1-2an_0.\label{Lutt}\ee
Here $a$ is the 1-dimensional scattering length. This is defined in terms of the 
1-dimensional even-channel phase shift, $\delta (k)$.  The even wave functions 
have the asymptotic long-distance behaviour:
\be \psi_e(x)\to \sin [k|x|-\delta (k)],\ee
where $x$ is the separation of the 2 bosons. 
At $k\to 0$ the phase shift is linear in $k$:
\be \delta (k)\to ak,\label{da}\ee
implying:
\be \psi_e(x)\to \sin k(|x|-a).\ee
Thus the relevance or irrelevance of pinning, in the dilute limit 
is determined by the sign of $a$. 
It is important to realize that this crucial sign is {\it not} 
fixed by the requirement that the boson-boson interaction be repulsive. 
For example, an infinite hard core repulsion of range $a_0$ leads 
to a scattering length $a=a_0>0$. On the other hand, a 
repulsive $\delta$-function interaction, $v\delta (x)$, 
leads to a {\it negative} scattering length, $a=-1/(\mu v)$ 
(where $\mu$ is the reduced mass). 
In a confined geometry the 1 dimensional scattering length 
depends not only on the direct inter-particle interaction but also 
on the effects of the boundaries. This problem was solved by Olshanii\cite{Olshanii} for
the case of ultra-cold atoms in a harmonic cylindrical trap, where 
it was shown that the sign of $a$ can be positive or negative 
depending on the ratio of the (positive) 3 dimensional scattering length 
to the trap radius. 

In this paper we study the properties of two interacting vortices in a thin 
platelet, or equivalently of 2 interacting bosons restricted to a narrow strip. 
We begin with the usual modified Bessel-function interaction between vortices 
given by anisotropic London theory. Standard boundary conditions at 
the edges of the platelet imply the existence of an infinite set of 
image vortices for each physical vortex. The interaction of 
an isolated vortex with its images, and with the external magnetic 
field determines its wave-function, $f (y)$, in the quantum mechanical analogue, determining 
the probability of the vortex being at a distance $y$ from the 
centre of the platelet. It is 
peaked near the centre, $y=0$. We then consider the scattering of two physical vortices, 
taking into account the interactions with all image vortices.  In general 
the two vortices could move off centre (away from $y_1=y_2=0$) as they scatter. 
Thus the calculation of the 
effective 1D scattering length requires, in principle, solving for a 2-dimensional 
2-body wave-function.

Fortunately, there is a very large dimensionless number that appears 
quite generally in the thermodynamics of vortices, and which simplifies 
our calculations considerably. Consider, for simplicity, a macroscopic 
isotropic London superconductor of penetration depth $\lambda$ and 
coherence length $\xi$. We approximate the Gibbs free energy 
for $N$ vortices as:
\be G\approx \int d\tau \left[\tilde \epsilon_1\sum_{i=1}^N\left({d\vec r_i\over d\tau }\right)^2
+{\phi_0^2\over 8\pi^2\lambda^2}\sum_{i<j}K_0(|\vec r_i(\tau )-\vec r_j(\tau )|/\lambda )\right]. 
\label{Gg}\ee
Here $\phi_0=hc/(2e)\approx 2\times 10^{-7}\hbox{G.-cm.}^2$ is the flux quantum and 
\be \tilde \epsilon_1\approx {\phi_0^2\over 16\pi^2\lambda^2}\ln (\lambda /\xi ),\ee
the tilt modulus, is simply the energy per unit length of the vortex.\cite{DeGennes} 
 $\tau$ is the spatial 
co-ordinate  along the field direction and $\vec r_i(\tau )$ describes the shape 
of the $i^{th}$ vortex. We have approximated the vortex-vortex interaction 
as only depending on the difference of the $\vec r_i$'s at the same value of $\tau$ and 
used the standard London model result for the interaction energy per unit length 
between straight parallel vortices,\cite{DeGennes} given by the modified Bessel function, $K_0$.  
(This needs to be cut off at short distances of order $r_{ij}\approx \xi$.)
We approximate the partition function by an integral over vortex paths, $\vec r_i(\tau )$, 
weighted by the Boltzmann factor, $\exp [-G/(k_BT)]$. 
By identifying $G/(k_BT)$ with $S/\hbar$ where $S$ is the classical action for 
$N$ interacting bosons, the classical partition function describing thermal 
fluctuations of vortices becomes equivalent to the Feynman path integral 
for interacting bosons. In this way various 
thermal properties of the vortex system can be conveniently obtained 
from the quantum system.\cite{Nelson,Fisher} 
 (We set $\hbar$ and $k_B=1$.) The corresponding Hamiltonian is:
\be H=-{1\over 2m}\sum_{i=1}^N\nabla_i^2+{\phi_0^2\over 8\pi^2\lambda^2T}
\sum_{i<j}K_0(|\vec r_i(\tau )-\vec r_j(\tau )|/\lambda ) \ee
with
\be m=\tilde \epsilon_1/T.\ee
(Strictly speaking, even after inserting appropriate factors of $\hbar$ and $k_B$, 
both terms in this Hamiltonian have dimensions of inverse length rather than 
energy. This can 
be traced back to the fact that in Eq. (\ref{Gg}) $\tau$ is a spatial coordinate 
in the classical model but is treated as an imaginary time in the analogue 
quantum one. This creates no problems for our analysis since physical 
quantities calculated using this quantum approach involve appropriate 
ratios of parameters with the correction dimensions, as we shall see.)
It is convenient to change to dimensionless length variables, letting:
\be \vec u_i\equiv \vec r_i/\lambda .\ee
We then may write the Hamiltonian in dimensionless form:
\be 2m\lambda^2H=-\sum_{i=1}^N\nabla_{i}^2+V_0\sum_{i<j}K_0(|\vec u_i-\vec u_j|)\ee
where the dimensionless parameter which measures the interaction strength is:
\be V_0={\tilde \epsilon_1\phi_0^2\over 4\pi^2T^2}\approx \left({\phi_0^2\over 8\pi^2T\lambda}\right)^2
\ln (\lambda /\xi).\ee
Noting that
\be \phi_0^2/(8\pi^2k_B)=3.9223\times 10^4\ \mu\hbox{m}\ -\ \hbox{K}\ee
we see that $V_0\gg 1$ for essentially 
any superconductor  at any $T<T_c$. This means 
that the  analogue quantum mechanical bosons have very strong short-range interactions when measured 
in dimensionless units. Variants of this large number will appear when 
we consider the potential energy function that holds the 
vortices in the middle of the slab and the interaction between vortices inside the slab. 
This implies that the vortices 
stay near the centre of the slab up to rather large slab widths 
justifying a 1 dimensional approximaton.  It also 
allows an unusual but powerful semi-classical approximation to be applied to the 1 dimensional problem
 yielding an explicit formula for the scattering length as a function 
the platelet thickness $d$ and other parameters ($\lambda_a$, $\lambda_c$, 
$\xi_a$, $\xi_c$ and $T$). Our conclusion is that 
 $a$ is positive and large for narrow platelets, increasing 
with $d$ and having a value $a\approx 19 \lambda_a$ for 
$d=10\lambda_c\approx .5 mm$.  This implies that columnar pins are relevant 
and also that the system rapidly leaves the dilute regime at 
low densities of order $1/(20\lambda_a)$, corresponding to fields, $H$ 
only slightly above $H_{c1}$.

In the next section we briefly review London theory and discuss properties of a single vortex in a thin 
platelet.  In Sec. III we consider 2 interacting vortices, determining 
the scattering length.  Sec. IV contains conclusions. 

\section{A single vortex in a thin platelet superconductor}

\begin{figure}
\centerline{\includegraphics[width=7.5cm,clip]{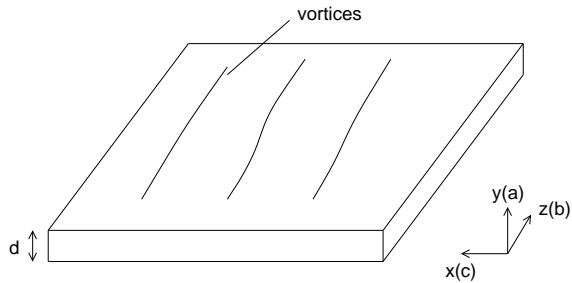}}
\caption{A thin platelet superconductor.}
\label{fig:slab}
\end{figure}

We make the London approximation,\cite{DeGennes} valid when the penetration depth is 
much longer than the coherence length, $\lambda \gg \xi$. We label 
the direction perpendicular to the platelet the $y$-direction, 
and label the direction of the magnetic field the $z$ direction. 
In a YBCO crystal the most promising geometry may be choosing $y$ and $z$ 
to be the $a$ and $b$ directions (or vice versa). See Fig. (\ref{fig:slab}). 
 Thus the thin direction of the platelet is the $a$ direction, not 
the usual growth direction, which is $c$. Such a sample could presumably 
be obtained by cleaving a macroscopic sample. The magnetic field of 
a single vortex, centered at $\vec r=(x,y)=0$ thus obeys:
\be h - \left[\lambda_a^2\frac{\partial^2 h}{\partial x^2} + \lambda_c^2 \frac{\partial^2 h}{\partial y^2} \right] = \phi_0\delta^2(r).
\label{london}\ee
 The Dirac $\delta$-function at 
the vortex core 
should actually be smeared over a distance of order $\xi$, the coherence length. 
Note that the decay of the magnetic field in the $x=c$ direction is governed 
by supercurrents running in the in the $y=a$ direction and hence involves $\lambda_a$ 
whereas the decay in the $y=a$ direction is governed by supercurrents 
running in the $x=c$ direction and hence involves $\lambda_c$.  In extremely 
underdoped YBCO crystals typical parameter values are 
\bea \lambda_c&=&50 \mu \text{m}\nn
\lambda_a&=& .5\mu \text{m}\nn
\xi_a &=& 5 \text{nm} \nn
\xi_c&=&.05 \text{nm}\nn
T_c&=& 17 K.
\label{par}\eea
(The value of $\xi_c$ may be a bit small compared to existing measurements but 
it is convenient to assume the result which follows from anisotropic 
Ginsburg-Landau theory: $\lambda_c/\lambda_a=\xi_a/\xi_c$. In any 
event our results only depend logarithmically on the $\xi$'s.)
Thus the vortex is extremely elliptical: much more extended in the 
$y=a$ direction. To acheive the two-dimensional limit, we need the 
sample thickness to be of order the vortex size.  (Actually, as we 
shall see a thickness of up to ten times the vortex size or more is alright.) 
Thus we can take advantage of the larger $\lambda_c$ by cleaving 
our crystal in the $a$-direction. We will refer to the parameters 
in Eq. (\ref{par}) at temperature $T\approx T_c$, with $a=y$ 
the thin direction as the standard parameters. However, our 
results should also apply at lower temperatures.  Note that 
the vortex wandering that we are concerned with here
 occurs primarily in the $x=c$ direction. The vortices presumably 
feel a periodic potential with wave-length given by 
the lattice constant in the $c$-direction. We will ignore this 
here.  In the dilute limit that we are considering it is 
not expected to have an important effect. 

The solution of Eq. (\ref{london}) 
at distances $r\gg \xi$, ignoring for now the boundaries, can be found by using a Fourier transform and is 
\bea h &=& \frac{\phi_0}{4\pi^2} \int_{-\infty}^\infty \int_{-\infty}^\infty \frac{e^{i(xk_x+yk_y)}}{\lambda_a^2k_x^2 + \lambda_c^2k_y^2 + 1}dk_xdk_y \\
&=& \frac{\phi_0}{2\pi\lambda_a\lambda_c}K_0\left(\sqrt{\left(\frac{x}{\lambda_a}\right)^2 + \left(\frac{y}{\lambda_c}\right)^2}\right)
\label{h}\eea
where $K_0$ is a modified Bessel function.

The energy per unit length of a vortex is 
\be \epsilon = -\frac{1}{8\pi}\int h \left[\lambda_a^2\frac{\partial h}{\partial x}\hat{x} + 
\lambda_c^2\frac{\partial h}{\partial y}\hat{y}\right] \cdot d\sigma, \ee
where the integral is taken over an ellipse of radii  $\xi_a$, $\xi_c$ around the vortex core. Thus 
\be \epsilon \approx \frac{\phi_0^2}{16\pi^2\lambda_a\lambda_c} \ln\left(\frac{\lambda_c}{\xi_a}\right)
, \label{line}\ee
to logarithmic accuracy. 
For the parameters in Eq. (\ref{par})  
we have $\epsilon = 10^{-8} \text{erg/cm}.$

We remark that the interaction energy per unit length between 2 straight parallel vortices separated by a vector $\vec r$
is simply:
\be U_{12}=\frac{\phi_0h_{12}}{4\pi},\ee
where $h_{12}(\vec r)$ is the magnetic field at the location of one vortex produced by the other, Eq. (\ref{h}).

We now consider a single straight vortex in an infinite slab, of thickness $d$, extending from $-d/2$ to $d/2$.
The presence of boundaries of the superconductor at $y=\pm d/2$, imposes boundary conditions, 
\be 0=j_y= {\partial h\over \partial x}.\ee
A vortex at position $y$ in a slab of thickness $d$ creates image vortices with magnetic field in the opposite 
direction at positions $(2n+1)d - y$ for integral $n$, and creates image vortices with field in the 
same direction at positions $2nd + y$ for all nonzero integral $n$. Thus the magnetic field at location $(x,y_2)$ 
for a vortex at $(0,y_1)$ is:
\bea h(x;y_2,y_1)&=& \frac{\phi_0}{2\pi\lambda_a\lambda_c}\left[\sum_{n=-\infty}^\infty K_0\left(\sqrt{\left(\frac{x}{\lambda_a}\right)^2
+\left(\frac{2nd + y_1-y_2}{\lambda_c}\right)^2}\right)\right.\nn  & & \left.- \sum_{n=-\infty}^\infty 
K_0\left(\sqrt{\left(\frac{x}{\lambda_a}\right)^2+\left(\frac{(2n+1)d - y_1-y_2}{\lambda_c}\right)^2}\right)\right]. \label{hI} \eea
In addition to the field of the vortex and its images an additional field occurs inside the superconductor when 
a field $H$ is applied outside of it:
\be h_1(y) = H\frac{\cosh(y/\lambda_c)}{\cosh(d/2\lambda_c)}. \label{h_1}\ee
The magnetic field at the position of the vortex, $(0,y)$ due to all its image vortices is:
\be h_2(y) = \frac{\phi_0}{2\pi\lambda_a\lambda_c}\left[- \sum_{n=\pm 1, \pm 3...} K_0\left(\frac{nd - 2y}{\lambda_c}\right) +
 \sum_{n=\pm 2, \pm 4...} K_0\left(\frac{nd}{\lambda_c}\right)\right].\label{h2}\ee
The Gibbs free energy depends on the position $y$ of the vortex as:
\be V_1(y) = \frac{\phi_0}{4\pi}[h_1(y) + \frac{1}{2}h_2(y)]+\hbox{constant}. \label{V_1}\ee

\begin{figure}[htp]
  \begin{center}
    \subfigure[Gibbs potential, $d = 10\lambda_c$] {\label{Vy} \includegraphics[width = 3.7cm, clip]{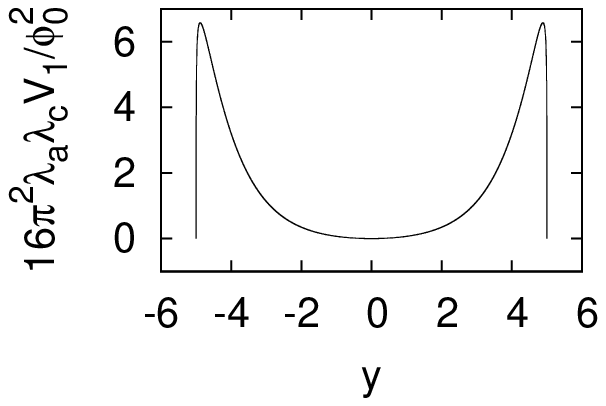}}
    \subfigure[Gibbs potential, $d = 100\lambda_c$] {\label{V100} \includegraphics[width = 3.7cm, clip]{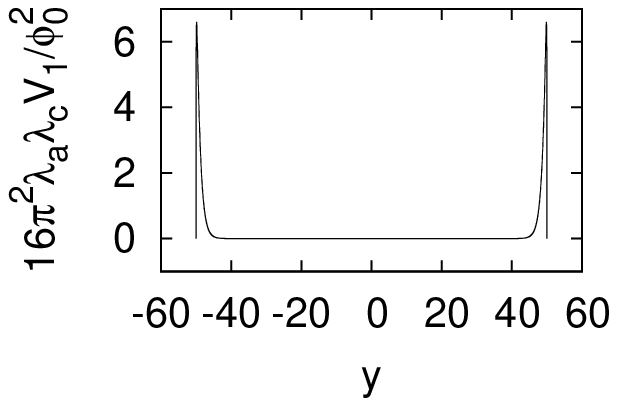}} \\
    \subfigure[Wave function, $d = 10\lambda_c$] {\label{psi10} \includegraphics[width = 3.7cm, clip]{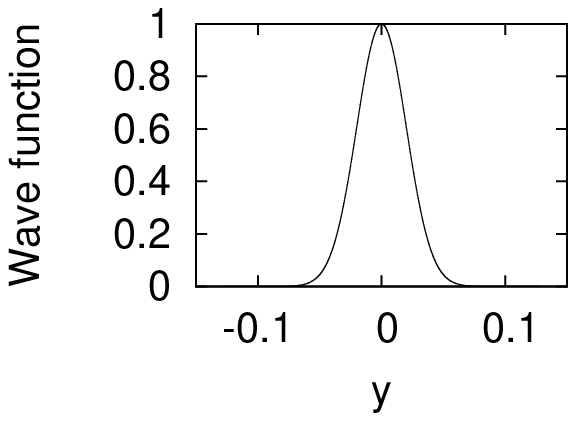}}
    \subfigure[Wave function, $d = 100\lambda_c$] {\label{psi100} \includegraphics[width = 3.7cm, clip]{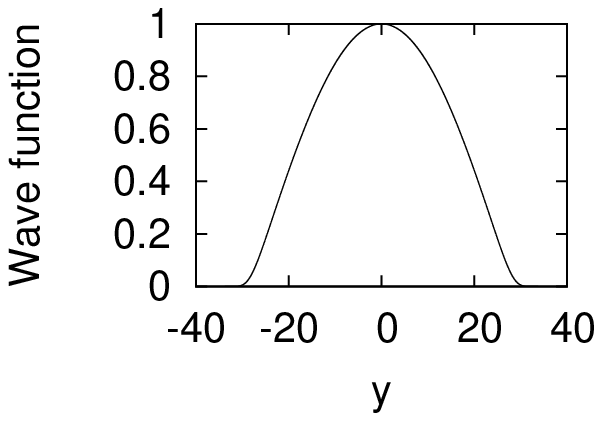}}
  \end{center}
  \caption{(a) and (b):The Gibbs potential,  Eqs. (\ref{h_1}) - (\ref{V_1}),
 for a single vortex with our standard parameters, $H=H_{c1}$ and $d = 10,\ 100\lambda_c$; 
 $y$ is in units of $\lambda_c$.
(c) and (d): The corresponding wave functions $f(y)$.}
  \label{fig:edge}
\end{figure}

This is plotted in Fig. (\ref{Vy}) and (\ref{V100} at $H=H_{c1}$.  It has a minimum at $y=0$, large barriers at intermediate $y$ 
and then appears to diverge to $-\infty$ at $y\to \pm d/2$:
\be V_1(y)\to -{\phi_0^2\over 16\pi^2\lambda_a\lambda_c}\ln [\lambda_c/(d\pm 2y)].\ee
 This divergence is due to 
the interaction of the vortex with its image at $y\mp d$. This divergence should actually 
be cut off at $y\mp d/2$ of order $\xi$ due to corrections to London theory. 
We take this into account by replacing $K_0[(nd-2y)/\lambda_c]$ by $K_0[(nd-2y)/\lambda_c]-
K_0[(nd-2y)/\xi ]$ in Eq. (\ref{h2}).
For $H>H_{c1}$,  we might expect the true minimum energy 
to be at $y=0$. This gives the usual formula for $H_{c1}$ for an anistropic superconductor:
\be H_{c1}\approx {4\pi \over \phi_0}\epsilon .\label{Hc1}\ee
where $\epsilon$ is the energy per unit length of a bulk vortex, Eq. (\ref{line}). 

We now turn to a study of thermal fluctuations of a single vortex inside the slab for 
such an anisotropic superconductor. With the magnetic field along the $b$-axis, the 
tilt modulus is very different for vortex tilting in the $a=y$ or $c=x$ direction. The 
elastic energy is written:
\be G_0=\int d\tau \left[{\tilde \epsilon_a\over 2}\left({dy\over d\tau }\right)^2+
{\tilde \epsilon_c\over 2}\left({dx\over d\tau }\right)^2\right] .\ee
Due to the assumed symmetry under rotations in the $a$-$b$ plane, the tilt 
modulus for tilting in the $a$ direction is just given by the energy per unit 
length:
\be \tilde \epsilon_a=\epsilon \ee
where $\epsilon$ is given in Eq. (\ref{line}). On the other hand, 
the energy per unit length of a vortex aligned parallel to the $c=x$ axis is much larger, 
resulting in the tilt modulus\cite{Ivley}
\be \tilde \epsilon_c={\phi_0^2\lambda_c\over 8\pi^2\lambda_a^3}\ln (\lambda_c/\xi)\approx 
{2\lambda_c^2\over \lambda_a^2}\tilde \epsilon_a.\label{epc}\ee
To this must be added the $y$-dependent free energy:
\be G_1=\int d\tau V_1[y(\tau )].\ee
Here we have again considered only ``instantaneous'' interactions between the vortex and 
its images, at a fixed value of $\tau$. 
We do not expect this approximation to qualitatively change the long distance physics in the dilute limit.  
  If we consider 
a very long vortex, in a sample of macroscopic length in the $z=\tau=b$ direction, then 
the probability of the displacement of the vortex from the centre of the slab having some value $y$ 
is simply given by $|f(y)|^2$ where $f$ is the ground state wave-function of the one-dimensional 
Hamiltonian:
\be H_1=-{1\over 2m_a}\left({d\over dy}\right)^2+{V_1(y)\over T}\ee
with
\be m_a\equiv \tilde {\epsilon_a\over T}.\label{ma}\ee
It is now convenient to define the dimensionless length variable:
\be \tilde y \equiv y/\lambda_c,\ee
in terms of which:
\be 2m_a\lambda_c^2H=-\left({d\over d\tilde y}\right)^2+V_{0y}\left[-\sum_nK_0[(2n+1)d/\lambda_c-2\tilde y]
+\ln (\lambda_c/\xi_a){\cosh (\tilde y/2)\over \cosh (d/2\lambda_c)}\right] \ee
where
\be V_{0y}\equiv {2\tilde \epsilon_a\phi_0^2\lambda_c\over 16\pi^2\lambda_aT^2}=
2\left({\phi_0^2\over 16\pi^2\lambda_aT}\right)^2\ln \left({\lambda_c\over \xi_a}\right). \label{V0y}
\ee
[We have set $H=H_{c1}$ given in Eq. (\ref{Hc1}).]
Since we will choose $d$ of order $\lambda_c$, we see that the dimensionless number $V_{0y}$ 
characterizes the height of the barriers holding the vortex at the center of the platelet. 
Using the numbers in Eq. (\ref{par})  and choosing $T=T_c$ we find:
\be V_{0y}\approx 10^8.\ee
We solve this Schroedinger equation numerically for the groundstate, for $d=10\lambda_c$, 
 Fig. [\ref{psi10}], finding that the particle makes only 
very small quantum fluctuations away from $y=0$ due to the huge barriers. 
For the above parameters we find 
\be \sqrt{<y^2>}=.01419\lambda_c=.001419 d.\ee
We should estimate the true critical field, $H_{c1}(d)$ 
using the ground state energy  of the quantum Hamiltonian. If the 
field is too low the particle can tunnel through the barrier corresponding 
to a vortex terminating at some value of $\tau$ with the magnetic flux 
leaving the superconductor. However, due to the exceedingly high barrier, 
the tunnelling probability is miniscule and the system will not 
be very sensitive to the precise value of $H$. 
We should also remark that, to solve the Schroedinger equation precisely we need 
to specifiy some boundary conditions on the wave-function at $y=\pm d/2$. 
We imposed vanishing boundary conditions. 
Fortunately, the extremely large dimensionless barrier also renders our 
results very insensitive to this choice. 

\section{two vortices}
Consider two straight parallel vortices inside the platelet, at locations $(x_i,y_i)$.  
The Gibbs free energy per unit length is:
\be V(x_1-x_2;y_1,y_2)={\phi_0\over 4\pi}h(x_1-x_2;y_1,y_2)+V_1(y_1)+V_2(y_2).\label{Vtot}\ee
By translational invariance in the $x$-direction, $V$ depends only on 
the difference of the $x$-coordinates of the two vortices:
\be x\equiv x_1-x_2.\ee
Here $h(x,y_1,y_2)$ is given in Eq. (\ref{hI}) and $V_1(y)$ by Eq. (\ref{V_1}). 
The first term in Eq. (\ref{Vtot}) represents the interaction of one 
vortex with the other one and with the images of the other one.  The second 
and third terms represent the interaction of each vortex with its own images 
and with the screened external field. 
$V(x,y_1,y_2)$ has a deep minimum at $y_1=y_2=0$ for large $|x|\gg \lambda_a$. $V(x,0,0)$ 
has a large peak centred at $x=0$ with a weak, logarithmic divergence right at $x=0$. 
This logarithmic divergence should be cut off at scales of order $\xi$; however 
this cut off has essentially no effect on the scattering length, as we shall see.   

Again we may study the thermodynamics of two wiggling vortices in the platelet by mapping 
onto a quantum mechanics model.  We make the fundamental assumption 
that the vortices only bend on long length scales (compared to $\lambda_a$)
and that we may approximate the vortex-vortex (and vortex-image vortex) interaction 
by an ``instantaneous'' one at a fixed value of $\tau =z$. 
The Boltzmann sum is now 
over the configuration of two vortices and we must include the vortex-vortex 
interaction in the free energy. Again identifying the free energy 
with the imaginary time action, we see that the corresponding quantum Hamiltonian is:
\be H = -{1\over 2}\sum_{i=1}^2\left[{1\over m_a}\left({d\over dy_i}\right)^2+
{1\over m_c}\left({d\over dx_i}\right)^2\right] +{V(x,y_1,y_2)\over T}.\label{SE2D}\ee
Here $m_a$ is given in Eq. (\ref{ma}) and
\be m_c\equiv {\tilde \epsilon_c\over T}\ee
where $\tilde \epsilon_c$ is given in Eq. (\ref{epc}). 
In the quantum analogue, the vortices obey Bose statistics\cite{Nelson} and consequently 
the two-body wave-function must be symmetric: even under $x\to -x$. The asymptotic 
behavior of the low energy wave-functions at $|x|\gg \lambda_a$ is given by:
\be \psi (x,y_1,y_2)\to f(y_1)f(y_2)\sin [k|x|-\delta (k)]\ee
where $f(y)$ is the ground state wave-function for a single vortex, discussed in Sec. II. 
The energy of this scattering state is:
\be E=2E_1+{k^2\over 2\mu}\ee
where $E_1$ is the ground state energy for a single vortex, discussed in the previous section 
and the reduced mass which governs the relative motion is
\be \mu = m_c/2.\ee
The scattering length is defined by Eq. (\ref{da}). 

It turns out that, due to the large barrier near $x=0$, for small $y_i$, the scattering length is 
determined almost completely by the large $x$ asymptotic form of the potential, until $d$ gets very large compared to $\lambda_c$. 
The $x$-dependent part of the exact potential can be written as a sum of exponentials:
\be V(x;y_1,y_2) = \frac{\phi_0^2}{4\pi\lambda_a^2d}\sum_{n=1}^{\infty} f\left( n,\frac{y_1}{d},\frac{y_2}{d}\right) 
\frac{e^{-|x|\sqrt{1+(n\pi\lambda_c/d)^2}/\lambda_a}}{\sqrt{1+(n\pi\lambda_c/d)^2}/\lambda_a} +V_1(y_1)+V_1(y_2).\ee
where we define
\be f(n,y_1/d,y_2/d) \equiv \left\{\begin{array}{rl} \cos(n\pi y_1/d)\cos(n\pi y_2/d) & 
\text{if $n$ is odd} \\ \sin(n\pi y_1/d)\sin(n\pi y_2/d) & \text{if $n$ is even} \end{array}. \right. \ee
At large $|x|$ we may approximate the sum by the first term only:
\be V \approx \frac{\phi_0^2\lambda }{4\pi\lambda_a^2d}e^{-|x|/\lambda}\cos(\frac{\pi y_1}{d})\cos(\frac{\pi y_2}{d})
+V_1(y_1)+V_1(y_2) \label{Vlx}\ee
where we have defined, for convenience, a reduced value of $\lambda_a$:
\be \lambda \equiv \lambda_a\left[1 + \left(\frac{\pi\lambda_c}{d}\right)^2\right]^{-1/2}. \label{lambda}\ee
For this approximation to hold, we need that the first term dominates all other terms. In the one-dimensional limit (i.e. $y_1=y_2=0$) the condition for large $x$ is
\be \exp\left(\frac{|x|}{\lambda_a}\frac{4\pi^2\lambda_c^2}{d^2}\right) \gg 1. \label{com} \ee
Note that, unlike the direct vortex-vortex interaction, this potential has a simple exponential 
dependence on $x$ at large $|x|$ albeit with a reduced penetration depth. 

Due to the large barriers in $V_1(y)$ we expect the low energy scattering states 
to be confined to $y_i\approx 0$. 
We first calculate $a$ assuming that the vortices stay exactly in the middle of the 
slab, $y_i=0$ throughout the scattering process. We return to a further discussion 
of why this is reasonable at the end of this section. 
This reduces the problem to a one-dimensional quantum mechanics model with Hamiltonian:
\be H=-{1\over 2\mu}{d^2\over dx^2}+{V(x)\over T}\ee
where
\be V(x)=V(x;0,0)\to \frac{\phi_0^2\lambda}{4\pi\lambda_a^2d}e^{-|x|/\lambda}.\ee
We look for parity even solutions of this Schroedinger equation with asymptotic 
behavior $\psi (x)\to \sin [k|x|-\delta (k)]$ with $\delta (k)\to ak$ as $k\to 0$. 
Note that in the small $k$ limit, $\psi (x)\to \sin [k(|x|-a)]$ for 
$x\gg \lambda$. Then, if we consider an intermediate range of $x$:
\be \lambda \ll |x|\ll 1/k,\ee
we may approximate the wave-function by a linear form:
\be \psi (x)\propto |x|-a.\label{x-a}\ee
Thus, to find the scattering length we need to simply solve the zero energy Schroedinger equation:
\be \left[ -{1\over 2\mu}{d^2\over dx^2}+{V(x)\over T}\right]\psi =0.\label{0}\ee
Note that this reduces the eigenvalue problem to 
a simple initial value problem.  We simply impose the initial conditions:
\bea {d\psi\over dx}(0)&=&0\nn
\psi (0)&=&1\eea
and solve the zero energy equation.
The asymptotic behavior of the solution at $|x|\gg \lambda$ is given by Eq. (\ref{x-a}) 
which determines the scattering length, $a$.

It is convenient to introduce a dimensionless length variable:
\be \tilde x\equiv x/\lambda \ee
in terms of which the Schroedinger equation becomes, at large $|\tilde x|$:
\be \left[ -{d^2\over d\tilde x^2}+V_{0x}e^{-\tilde x}\right]\psi (\tilde x)=2\mu \lambda^2E\psi (\tilde x).\label{Su}\ee
Here the dimensionless number which measures the strength of the repulsive potential is:
\be V_{0x}={\tilde \epsilon_c \phi_0^2\lambda^3\over 4\pi \lambda_a^2dT^2}=
{1\over 2\pi}\left({\phi_0^2\over 4\pi\lambda_aT}\right)^2
\ln (\lambda_c/\xi_a){\lambda_c/d\over [1+(\pi \lambda_c/d)^2]^{3/2}}. \label{V0x}\ee
Using our estimates of the parameters in Eq. (\ref{par}) with $T=T_c$ and $d=10\lambda_c$ we find:
\be V_{0x}=1.07 \times 10^8.\ee
Note that $V_{0x}\propto 1/d$ at $d\gg \lambda_c$. Importantly $V_{0x}\gg 1$ when 
$d$ is of order $\lambda_c$ and remains large out to extremely large 
values of $d/\lambda_c$. The largeness of $V_{0x}$ leads to a large scattering length, allows 
for an unusual semi-classical solution approximation and also helps to justify setting $y_i=0$ as we shall see below. 
It is of course, possible to solve the Schroedinger equation numerically 
for specified values of the parameters.  However, the largeness of $V_{0x}$ and 
$V_{0y}$ creates numerical difficulties for standard algorithms, when 
one attempts to solve the full 3-dimensional problem, including the $y_i$. 
In this case it is much easier, and more transparent, to use the semi-classical approximation. 

Our semi-classical approximation for the scattering length at $V_{0x}\gg 1$ begins with the observation that the classical 
turning point for 2 particles approaching each other with a small relative momentum, $k$ 
occurs at $\tilde x\gg 1$. Therefore $a$ is determined almost completely by the large $\tilde x$ 
form of the potential in Eq. (\ref{Vlx}).
Using the large $\tilde x$ form of the potential, we may 
solve the one-dimensional Schroedinger equation exactly. To do this we change variables to:
\be u = 2\sqrt{V_{0x}}e^{-|\tilde x|/2}.\ee
The zero energy Schroedinger equation, (\ref{0}), simplifies to
\be u^2\psi'' + u\psi' - u^2\psi = 0 \ee
where the primes denote differentiation with respect to $u$. This is the zeroth order modified Bessel's differential equation, and 
the general solution is given in terms of the modified Bessel functions:
\bea \psi &=& c_1 I_0(u) + c_2 K_0(u)  \nn
&=& c_1 I_0(2\sqrt{V_{0x}}e^{-|x|/2\lambda}) + c_2K_0(2\sqrt{V_{0x}}e^{-|x|/2\lambda}). \eea
For large $u$ (i.e. $\tilde x << \ln V_0$) we have
\be \psi \approx \frac{1}{\sqrt{2\pi u}} \left(c_1 e^u + c_2 \pi e^{-u} \right), \ee
or writing in terms of the original variables and putting back factors of $\lambda$, we have
\be \psi \approx \frac{1}{\sqrt{4\pi}V_{0x}^{1/4}e^{-|x|/4\lambda}} 
\left(c_1 e^{2\sqrt{V_{0x}}e^{-|x|/2\lambda}} + c_2 \pi e^{-2\sqrt{V_{0x}}e^{-|x|/2\lambda}} \right) \label{large} \ee
On the other hand, for small $u$ (i.e. $\tilde x >> \ln V_0$) we have 
\be \psi \approx c_1 - c_2 (\ln(u/2) + \gamma) = c_1 + c_2\left(\frac{|\tilde x|}{2} - \frac{1}{2}\ln V_{0x} - \gamma\right) \ee
where $\gamma \approx 0.5772$ is the Euler-Mascheroni constant. Note that the last formula can be written as
\be \psi \approx \frac{c_2}{2\lambda} (|x| - a) \ee
where the scattering length $a$ is (putting back factors of $\lambda$)
\be a = \lambda (\ln V_{0x} + 2\gamma - 2c_1/c_2). \label{scatteringlength} \ee
To determine $c_1/c_2$ we match our solution in the region $1\ll |\tilde x|\ll \ln V_{0x}$ to the WKB solution which 
 works for $|x|$ not too large, in the region where $V(x)$ is large. 
The even WKB wave function for $E = 0$ is given by \be \psi(x) = A \left\{ \exp\left[\int_0^{x} \sqrt{V'(x')} dx'
\right] + 
\exp\left[-\int_0^{x} \sqrt{V'(x')} dx'\right] \right\}\ee
where
\be V'(x)=2\mu V(x;0,0)\ee
and $V$ is the exact potential of Eq. (\ref{Vtot}). (Note that we {\it do not} make any large $x$ approximation 
to $V$ now.) We can rewrite this as
\be \psi(x) = A \left\{ e^{-\alpha} \exp\left[\int_x^\infty \sqrt{V'(x')} dx'\right] + 
e^\alpha \exp\left[-\int_x^\infty \sqrt{V'(x')} dx'\right] \right\} \ee
where 
\be \alpha = \int_0^\infty \sqrt{V'(x)} dx. \ee
Now, if $x$ is large enough for our asymptotic expression $V'(x) \approx V_{0x}e^{-x/\lambda}/\lambda^2$ 
 to hold, 
then the integral can be done quite readily:
\be \psi(x) = A \left\{ e^{-\alpha} \exp\left[2\sqrt{V_{0x}}e^{-x/2\lambda}\right] + e^\alpha \exp\left[-2\sqrt{V_{0x}}
e^{-x/2\lambda} \right] \right\}. \ee
Comparison with Eq. (\ref{large}) thus gives
\be c_1/c_2 = \pi e^{-2\alpha} = \pi \exp\left[-\int_{-\infty}^\infty \sqrt{V'(x)}dx\right].
\label{c1c2} \ee
This quantity is exponentially small in the large quantity $V_{0x}$ so it is completely negligible. 
Note also that the logarithmic divergence of $V(x;0,0)$ at $x\to 0$ has no important effects, 
leaving the integral finite in Eq. (\ref{c1c2}). 
 The last term in Eq. (\ref{scatteringlength})essentially  vanishes, and we simply have that
\be a = \lambda (\ln V_{0x} + 2\gamma). \label{a}\ee
Asymptotically, the scattering length is linearly dependent on the logarithm of the size of the potential.

Interestingly, we get almost the same result for the odd wave functions, except that the sign of $c_1/c_2$ is reversed. 
The even channel and odd channel scattering lengths are therefore almost exactly the same. However, it is the \emph{difference}
 between the even and odd channel scattering lengths that determines the transmission coefficient, and it is only then 
that $c_1/c_2$ plays an important role.

We have based this approximation on the assumption that there exists a region of separation $x$, such that the approximation
 Eq. (\ref{large}) holds, and the WKB approximation to the wave function also holds. Typically the matching is done around 
the region $x \approx a$; 
therefore we need that (using Eq. (\ref{a}) and Eq. (\ref{com}), and noticing 
$\lambda \approx \lambda_a$ for large $d$)
\be \exp\left[\left(\frac{2\pi\lambda_c}{d}\right)^2\ln (e^{2\gamma} V_{0x})\right] \gg 1 \label{wkbcond}\ee
This is the condition that must be satisfied for the formula Eq. (\ref{a}) to hold.
\begin{figure}
\centerline{\includegraphics[width=7.5cm,clip]{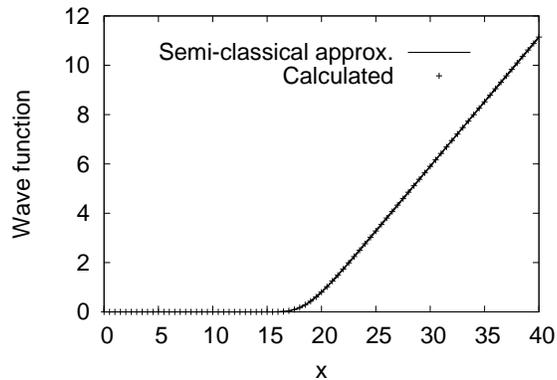}}
\caption{The  wave functions in the $k\to 0$ limit comparing a precise numerical 
calculation (in the 1 dimensional approximation) to the semi-classical approximation 
 for our standard parameters and $d=10\lambda_c$. Lengths are in units of $\lambda_a$.}
\label{1Dwave}
\end{figure}

\begin{figure}
\centerline{\includegraphics[width=7.5cm,clip]{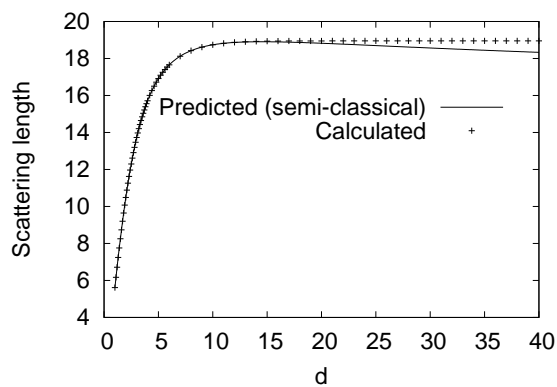}}
\caption{The scattering lengths based on a precise numerical 
calculation (in the 1 dimensional approximation) compared to our 
semi-classical approximation for our standard parameters.
 $d$ is in units of $\lambda_c$, the scattering length in units of $\lambda_a$.}
\label{fig:a}
\end{figure}

The wave-function calculated numerically to high precision (in the 1-dimensional approximation) 
and the semi-classical wave function are compared in Fig. (\ref{1Dwave})
for our standard parameters and $d=10\lambda_a$.As can be seen, the two wave functions give good agreement in the large-$x$ regime (with an error $< 1\%$ for $x > 18\lambda_a$). 
The semi-classical wave function is grossly inaccurate in the small-$x$ region ($x < 15\lambda_a$) but 
the wave-function is neglegible there anyway. 
The predicted semi-classical scattering length is $a = \lambda (\ln V_0 + 2\gamma) = 18.7398\lambda_a$. 
The actual scattering length of the numerically determined wave function (obtained by fitting the wave function in the large-$x$ regime to 
a linear function) is $18.7409\lambda_a$. The semi-classical wave function gives an error of less than $0.01\%$. 
In Fig. (\ref{fig:a}) we show the scattering length versus $d$, comparing our numerical results  to the semi-classical approximation
 (in both cases making the 1-dimensional approximation). 
 Most of the $d$-dependence in our semi-classical 
formula, Eq. (\ref{a}), arises from the $d$-dependence of the reduced penetration depth, $\lambda$, given 
in Eq. (\ref{lambda}). As the sample thickness decreases, the effective range of the interaction potential, $\lambda$,  
also decreases,  and the scattering length simply scales with it, up to logarithmic corrections coming 
from $V_{0x}$, defined in Eq. (\ref{V0x}). 
The semi-classical and numerical
 values agree within $1\%$ up to about $d = 22 \lambda_c$. For $d = 22 \lambda_c$, we have that (refer to Condition (\ref{wkbcond}))
\be \left(\frac{2\pi\lambda_c}{d}\right)^2\ln (e^{2\gamma} V_{0x}) = 1.56 \ee
which is already not far from unity. There will therefore be significant deviations of the true value from Eq. (\ref{a}).
 We could also notice that the scattering length tends towards a finite value for large $d$. This is because as the thickness of the
 sample grows, the image vortices move farther away from the orginal vortices until their effects become negligible.

\begin{figure}
\centerline{\includegraphics[width=7.5cm,clip]{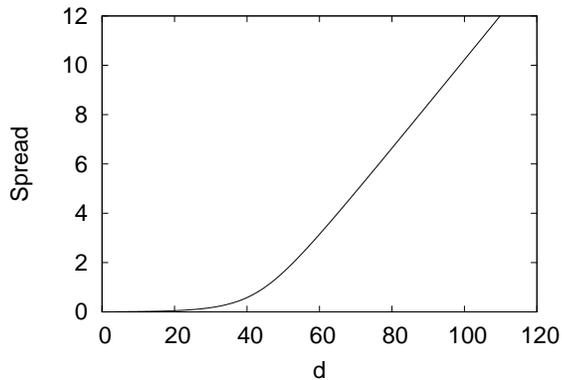}}
\caption{The spread of the single-vortex wave function, $\sqrt{<y^2>}$, plotted against $d$ using standard parameters.
 Both lengths are in units of $\lambda_c$.}
\label{width}
\end{figure}

Next, we discuss the validity of our 1-dimensional approximation, setting $y_i=0$, which is 
justified by the fact that the single vortex wave-function, $f(y)$ is so sharply peaked near $y=0$. 
If we look at the shape of the potentials, $V_1(y)$, 
we see that for small $d$ (Fig. (\ref{Vy})) the potential has an obvious minimum in the center, and is approximately
 simple harmonic near the center. For larger $d$ (Fig. (\ref{V100})), however, the potential is almost negligible 
except for a large potential barrier close to (but not at) the edges; the potential will be qualitatively more 
similar to a square well. Therefore as $d$ increases, we would expect the shape of the wave function to morph 
from a confined Gaussian to a spread-out sinusoidal (Fig. (\ref{psi10}), Fig. (\ref{psi100})).
The spread of the single-vortex wave function ($\langle y^2 \rangle ^{1/2}$) is plotted in Fig. (\ref{width}). 
The thickness reaches $1\%$ of the platelet thickness at around $d = 35\lambda_c$, an indication that our one-dimensional 
approximation fails above this value. The fact that the thickness becomes linearly dependent on $d$ at 
large $d$ also suggests that the wave function tends to a fixed shape (a sinusoidal).

 A more systematic 
approximation to solving the full 2-body 2-dimensional Schroedinger equation of Eq. (\ref{SE2D}), would be to write:
\be \psi (x;y_1,y_2)\approx \psi_1(x)f(y_1)f(y_2)\ee
where $\psi_1(x)$ is the 1-dimensional wave-function found above 
and $f(y_1)$ is the single vortex wave-function. We could then improve our estimate 
of the 1-dimensional effective potential by using:
\be V(x)\approx \int dy_1dy_2|f(y_1)|^2|f(y_2)|^2V(x;y_1,y_2)\ee
rather than simply $V(x)\approx V(x;0,0)$. However, because 
$f(y)$ is so sharply peaked at $y\approx 0$ this makes a negligible difference.

Note that our calculation of $a$ depended essentially only on $V(x)$ in the 
large $x$ region, $x\gg \lambda$. Our consideration of the small $x$ region 
only served to determined $c_1/c_2$ which was exponentially small anyway and can 
simply be ignored. In this large $x$ region, Eq. (\ref{Vlx}) is 
a good approximation, the wave-function approximately factorizes and the large barriers in $V_1(y)$ ensure 
that the wave-function is strongly peaked near $y_i=0$. For smaller 
values of $x$ the wave-function presumably spreads out more in the 
$y$ direction. However, at smaller $x$ the wave-function is exponentially small 
anyway.

Finally, we consider the case where the thin direction of the 
YBCO platelet is the $c$ direction: y=c (and x=a, z=b). In this case the roles of $\lambda_a$ 
and $\lambda_c$ are switched, as are the roles of $\tilde \epsilon_a$ 
and $\tilde \epsilon_c$ and we get 
\bea V_{0y} &=&2{\tilde \epsilon_c\phi_0^2\lambda_a\over 16\pi^2T^2\lambda_c}=
4\left({\phi_0^2\over 16\pi^2\lambda_aT}\right)^2\ln (\lambda_c/\xi_a) \nn
 V_{0x}&=&{\tilde \epsilon_a \phi_0^2\lambda^3\over 4\pi \lambda_c^2dT^2}=
{1\over 4\pi}\left({\phi_0^2\over 4\pi\lambda_aT}\right)^2
\ln (\lambda_c/\xi_a){\lambda_a/d\over [1+(\pi \lambda_a/d)^2]^{3/2}} \eea
where we now define:
\be \lambda \equiv\lambda_c\left[1+\left({\pi \lambda_a\over d}\right)^2\right]^{-1/2}.\ee
Apart from some unimportant factors of $2$, the formulas for $V_{0y}$ and $V_{0x}$ are the same as 
for the other geometry except that $d$ now appears in the dimensionless ratio 
$\lambda_a/d$ rather than $\lambda_c/d$. So we now conclude that 
the 1 dimensional approximation holds for $d\leq 20\lambda_a\approx 10 \mu$m. 
and the semi-classical approximation holds out to roughly the same value of $d$. 

\section{conclusions}
Our main result is the formula Eq. (\ref{a}) for the scattering length.  This is 
plotted versus the platelet thickness, $d$, for our standard parmeters, 
in Fig. (\ref{fig:a}). Note that $a$ is everywhere positive and $a/\lambda_a$ is 
everywhere quite large, having the value $a/\lambda_a\approx 19$ at $d\approx 10\lambda_c$.
At somewhat larger values of $d$ we expect our semi-classical approximation to 
the one-dimensional problem to break down and, more problematically, the 
one-dimensional approximation itself to start to fail. 

The main use of our formula for $a$ is not, of course, for studying the 
system with only 2 vortices, but rather for studying the thermodynamic 
limit of many vortices. In the dilute limit, $n_0a\ll 1$ (where $n_0$ is 
the vortex density per unit length),  $a$ determines the 
Luttinger parameter via Eq. (\ref{Lutt}). Of course the Luttinger liquid 
treatment of the problem assumes that it is fundamentally 1-dimensional. 
Our calculations here indicate that the 1-dimensional approximation should 
be good, at least in the 
dilute limit $n_0a\ll 1$, up to platelet thicknesses of order $d=10\lambda_c$ or more, 
 since the vortices stay very close to 
the centre of the platelet. Furthermore, we have determined the Luttinger 
parameter for this range of thicknesses and vortex densities. 
When $a$ is large the Luttinger parameter decreases rapidly for 
increasing vortex density. It was argued in (\onlinecite{Affleck}) that, 
at high densites, $g\ll 1$. Taken together, these results suggest a rapid monotonic drop 
of $g$ from $1$ with increasing density. In this case, columnar pins 
would be highly relevant for essentially all fields above $H_{c1}$. 
  Thus 
a promising region to look at experimentally might be very close to $H_{c1}$ with 
low vortex densities, $n_0\ll 1/\lambda_a$ and samples of 
thickness around $10\lambda_c$.
\acknowledgements 
We  would like to thank Doug Bonn, 
David Broun, David Nelson and Eran Sela for many helpful discussions. 
IA thanks Matt Choptuik and Brian Martin  for their collaboration in 
an earlier attack on this problem. This research is supported 
in part by NSERC (CL and IA) and CIfAR (IA).

\end{document}